\documentclass[twocolumn,aps,pra,groupedaddress,showpacs]{revtex4}
\usepackage{epsfig,amssymb}  

\begin{document}

\title{Separability conditions from entropic uncertainty relations}

\author{Vittorio Giovannetti}
 \altaffiliation[Recently at: ]{NEST-INFM \& 
Scuola Normale Superiore, Piazza dei Cavalieri 7, I-56126, Pisa, Italy.}
\affiliation{Research Laboratory of Electronics,
MIT - Cambridge, MA 02139, USA}

\begin{abstract}
We derive a collection of separability conditions for bipartite 
systems of dimension $d\times d$
 which is based on the entropic version of the 
uncertainty relations. A detailed analysis of the two-qubit case is
given by comparing the new separability conditions with existing criteria.
\end{abstract}


\maketitle

Separability is the property that distinguishes 
statistical ensembles that have a classical description
from the ones that need a quantum description.
As a matter of fact, the definition of
entanglement is the formal negation of the
separability condition \cite{CHU}. 
In spite of this clear logical distinction 
between separable and entangled states, the borderline
between these two sets is very difficult to 
characterize in practice. 
In the case of bipartite systems, many necessary
conditions for separability (i.e. criteria for entanglement) 
have been proposed
 \cite{PER,WIT,EKE,SIM,DUA,REID1,CAM,KOR,HOF}. 
Some of these conditions 
pertain to geometrical properties
of separable states which are difficult to
observe experimentally \cite{PER}. 
The ones that have a better chance to become operative
procedures for testing the presence of entanglement are 
those criteria that establish boundaries either on the
expectation values \cite{WIT,EKE}, or on the statistical variances  
\cite{SIM,DUA,REID1,CAM,KOR,HOF}
of some observables.
These last separability conditions 
take advantage of the fact that, when specified on nonentangled
states, the uncertainty relations of a collection of nonlocal observables 
$\hat{X}_1, \hat{X}_2, \ldots$ (i.e., observables that act nontrivially in 
both the Hilbert spaces which constitute the bipartite system) 
are forced to satisfy lower bounds, which are higher 
than the ones they have to  
obey when applied to generic states \cite{SIM,DUA,REID1,CAM,KOR,HOF}. 
The reason for this is that, in general, 
$\hat{X}_1, \hat{X}_2,\ldots$ 
possess only entangled eigenstates: on such states 
these observables are allowed to achieve the minimum values 
of their uncertainties.
On the other hand, because no 
common eigenstate of $\hat{X}_1, \hat{X}_2, \ldots$ is separable, 
it is not possible to minimize the uncertainties of all these observables 
simultaneously on such configurations.

In this paper we propose an approach to the separability 
problem of a bipartite system based on the entropic 
uncertainty relations in place of the usual Heisenberg-like uncertainty
relations. A somehow analogous endeavor has been undertaken in 
\cite{ADA} where the standard Bell inequalities were written
in terms of entropic quantities.
The strategy we propose
takes direct advantage of the geometrical structure of the tensor product
Hilbert space of the system and underlines the connections
between uncertainty relations and entanglement \cite{CER}.
The material is organized as follows. In Sec. \ref{SezEnt}, 
we give a brief
review of the entropic relations. In Sec. \ref{SezSep}, by analyzing the
simple case of a two-qubit system, we show how it is possible
to generate a new class of separability conditions using
entropic uncertainty relations of nonlocal operators.
In Sec. \ref{s:Phigh}, we generalize these results to 
bipartite systems of dimension $d\times d$ with $d\geqslant2$.


\section{Entropic uncertainty relations} \label{SezEnt}
Entropic 
uncertainty relations have been proposed as an
alternative to the standard Heisenberg-like relations, in the
case of observables with nontrivial C-number commutators
\cite{CER,DEU,MAJ,MAA,SAN,KRI}. The basic idea of this approach is
to replace the statistical variance with the Shannon entropy  
as an estimator of the uncertainties associated 
with the measurement process.
Consider for instance an observable $\hat X$ with $K$ distinct
eigenvalues $x_{1}, \ldots x_{K}$ and spectral decomposition
\begin{eqnarray}
    \hat X = \sum_{k=1}^{K} x_{k} {\mathbb X}_{k}
\label{spettro} \;,
\end{eqnarray}
with  ${\mathbb X}_{k}$ the projector in the eigenspace of $\hat X$
relative to the eigenvalue $x_{k}$.
Given a state $\rho$ of the system, we define the entropic
uncertainty of $\hat X$ as
\begin{eqnarray}
    H({\mathbb X}, \rho)  \equiv -\sum_{k=1}^{K} P_{k} \ln P_{k}
    \label{entro}
\end{eqnarray}
where $P_{k}\equiv\mbox{Tr}( {\mathbb X}_{k} \rho)$
is the  probability
of finding the state $\rho$ in the $k$th eigenspace.
In Eq.~(\ref{entro}) the symbol
$H({\mathbb X}, \rho)$ is used to underline the
dependence of this quantity from
 the projectors ${\mathbb X}_k$ 
defined in Eq.~(\ref{spettro}): this notation allows us 
to distinguish the
definition of entropic uncertainty of
$\hat X$ introduced here 
from the standard definition $H(\hat X, \rho)$
adopted elsewhere \cite{NOTA1}.
$H({\mathbb X}, \rho)$ can be used to estimate the  
uncertainty of the outcome of a measurement of $\hat X$
on the state $\rho$.
In fact, if $\rho$ is one of the eigenvectors belonging, say,
to the $k_{0}$th eigenspace, then $P_{k} = \delta_{k,k_{0}}$ and
$H({\mathbb X}, \rho)$ nullifies. On the contrary, if  $\rho$ is
an equally weighted 
superposition or mixture of 
all  the eigenstates of $\hat X$, the measurement result is 
maximally undetermined and $H({\mathbb X}, \rho)$ achieves
its maximum value $\ln K$. 
Now introduce a second observable $\hat Y$ with 
eigenspaces projectors $\{{\mathbb Y}_{k'}\}$. Following the
derivation of \cite{KRI}, it is straightforward to show that 
an entropic uncertainty relation applies, i.e.
\begin{eqnarray}
H({\mathbb X}, \rho) + H({\mathbb Y}, \rho)  \geqslant  -2 \ln 
\left( \max_{k,k'} 
||{\mathbb X}_{k} {\mathbb Y}_{k'} || \right) \; ,
\label{entropico}
\end{eqnarray}
where $H({\mathbb Y}, \rho)$  is the entropic uncertainty of $\hat Y$ 
and  where $|| {\cal O} || \equiv \max_{|\psi\rangle} 
|| {\cal O}  |\psi \rangle || $
is the norm of the operator ${\cal O}$. 
In our approach Eq. (\ref{entropico}) replaces the
standard uncertainty relation which involves the product of the
statistical variances of the two operators. These two relations
are not completely equivalent, but both predict that
when $\hat X$ and $\hat Y$ commute, no constraint is imposed on
the accuracy with which we can measure them on the same state  (see for
example \cite{MAJ,MAA}).


\section{Separability conditions for qubits}\label{SezSep}

A separable state of a bipartite system ${\cal S}$ 
composed of 
subsystems $\cal A$ and $\cal B$ is any density matrix
$\rho_{sep}$ that can be expressed as a convex combination of 
tensor product states, as
\begin{eqnarray}
    {\rho}_{sep}=\sum_{n} \lambda_{n} \;
    |\psi_{n}\rangle_{a}\langle \psi_{n}|  
    \otimes |\phi_{n}\rangle_{b}\langle \phi_{n}| \,,
    \label{sepone} 
\end{eqnarray}
with $|\psi_{n}\rangle_{a}$ and $|\phi_{n}\rangle_{b}$ pure states
of the subsystem $\cal A$ and $\cal B$ respectively, and
$\lambda_{n} \geqslant 0$, $\sum_{n} \lambda_{n} =1$. Any state
of this form is, by definition, not entangled \cite{CHU}.
The aim of this paper is to give a class of entropic 
relations, such as Eq.~(\ref{entropico}), that can be violated by
entangled states but not by the states 
$\rho_{sep}$. We begin by considering the simple case where
$\cal A$ and $\cal B$ are both qubits:
the method will be then extended to subsystems of higher dimension in
Sec. \ref{s:Phigh}.


\subsection{First example}\label{SezJointA}

Consider the following observables 
\begin{eqnarray}
\hat X \equiv \sigma^{(1)}_{a}\otimes \sigma^{(1)}_{b} \;, \quad
\hat Y \equiv \sigma^{(2)}_{a}\otimes \sigma^{(2)}_{b} 
\;, \label{es1}
\end{eqnarray}
where $\sigma^{(j)}_{s}$, for $j=1,2,3$ and $s=a,b$, are the
Pauli operators acting
on the $s$ qubit. Because $\hat X$ and $\hat Y$ commute, 
the right-hand-side of
Eq. (\ref{entropico}) vanishes and no lower bound is required to
the sum of the entropic uncertainties of these operators.
For example, one can nullify both 
$H({\mathbb X}, \rho)$ and $H({\mathbb Y}, \rho)$ by choosing 
$\rho$ to be one of the four maximally entangled elements of the Bell basis
\begin{eqnarray}
|\Psi_{1}\rangle&=&(|00\rangle+|11\rangle)/\sqrt{2}\nonumber\\
|\Psi_{2}\rangle&=&(|00\rangle-|11\rangle)/\sqrt{2}\nonumber\\
|\Psi_{3}\rangle&=&(|01\rangle+|10\rangle)/\sqrt{2}\nonumber\\
|\Psi_{4}\rangle&=&(|01\rangle-|10\rangle)/\sqrt{2}\label{PBELL}\;,
\end{eqnarray}
where, for instance, $|01\rangle$ is the state 
$|0\rangle_a \otimes |1\rangle_b$
with $|0\rangle$ and $|1\rangle$  being the eigenvectors of the
the Pauli operator $\sigma^{(1)}$ relative to the eigenvalues
$+1$ and $-1$, respectively. 
On the other hand, if we consider separable states $\rho_{sep}$, 
it is possible to show that the
following inequality applies, 
\begin{eqnarray}
H({\mathbb X}, \rho_{sep}) + H({\mathbb Y}, \rho_{sep})  \geqslant  \ln 2 \; .
\label{sepent1}
\end{eqnarray}
This relation can be used to test the presence of
entanglement in the system: if some state violates
it, such a state cannot be separable.
Since the Shannon entropy is a concave function \cite{COVER}, 
to prove Eq. (\ref{sepent1})
it is sufficient to show that it applies to any pure separable state
\cite{MAJ,MAA,KRI}
\begin{eqnarray}
|\Psi_{sep}\rangle = |\psi\rangle_{a} \otimes|\phi\rangle_{b}\;,
\label{PSEP}
\end{eqnarray}
with
\begin{eqnarray}
|\psi \rangle_{a} &=& \cos\alpha
|0\rangle_{a} +e^{i 
\delta} \sin\alpha  |1\rangle_{a} \nonumber \\
|\phi \rangle_{b} &=& \cos\beta
|0\rangle_{b} +  e^{i 
\gamma} \sin\beta |1\rangle_{b} \; ,
 \label{vittorio}
\end{eqnarray}
where $\alpha$, $\beta$, $\delta$ and $\gamma$ are real parameters.
The observable ${\hat X}$ has the eigenvalues
$+1$ and $-1$, which are both two-time degenerate and have
eigenspaces generated by the vectors  $\{ | 00 \rangle, | 11 \rangle \}$ 
and $\{ | 01 \rangle, | 10 \rangle \}$, respectively.
The probabilities of
finding the state $|\Psi_{sep}\rangle$ in these eigenspaces 
can be then expressed as, 
\begin{eqnarray}
P_{+} &=& |\cos\alpha\cos\beta|^{2} +
|\sin\alpha \sin\beta|^{2}  \label{probA} \\
P_{-} &=& 
1-P_{+}  \nonumber
\;.
\end{eqnarray}
Consequently the entropic uncertainty of $\hat X$ is 
$H({\mathbb X}, |\Psi_{sep}\rangle)={\cal H}_2(P_{+})$
\begin{figure}[t]
\begin{center}
\epsfxsize=.6\hsize\leavevmode\epsffile{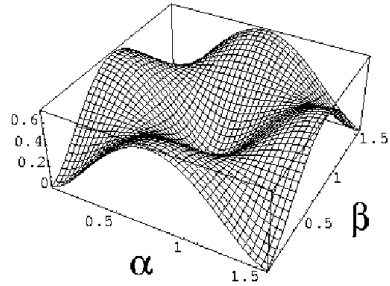}
\end{center}
\caption{Entropic uncertainty relations for the pure separable state
 $|\Psi_{sep}\rangle$ of Eq. (\ref{PSEP}). 
The plot shows the difference between 
the right-hand-side of Eq.~(\ref{sepentro1}) evaluated 
for $\delta=-\gamma=\pi/2$ and the
lower bound $\ln2$: this function is never negative.}
\label{fig1}
\end{figure}
with 
\begin{eqnarray}
{\cal H}_2(x)\equiv-x\ln x -(1-x) \ln(1-x)\;,
\label{PBINARY}
\end{eqnarray}
the binary entropy function.
In the same way we can calculate the entropic uncertainty of the 
operator $\hat Y$ and show that the following relation applies,
\begin{eqnarray}
&H({\mathbb X}, |\Psi_{sep}\rangle) + H({\mathbb Y}, |\Psi_{sep}\rangle)
=& \nonumber\\
&{\cal H}_2(|\cos\alpha \cos\beta|^{2} +
|\sin\alpha \sin\beta|^{2})&  \label{sepentro1} 
\\ &+ 
{\cal H}_2[ (1- \sin\delta\sin\gamma \sin(2\alpha) \sin(2\beta) ) /2]&\!\!\!.
\nonumber
\end{eqnarray}
We are interested in the minimum value achievable by this
four-parameter function. The analysis is simplified by
the fact that ${\cal H}_2(x)$ is a decreasing
function of $|1-2x|$. Hence, for any  
$\alpha, \;\beta \in [0,\pi/2]$ 
the right-hand-side of Eq. (\ref{sepentro1}) reaches its minimum for
\mbox{$\delta,\gamma=\pm \pi/2$}. Assigning these values for $\delta$ 
and $\gamma$, Eq. (\ref{sepentro1}) can then
be shown to have minimum equal to $\ln 2$  (e.g. see Fig.~\ref{fig1}),
concluding the proof \cite{NOTTATA}.


Entropic uncertainty relations can be derived for more than
two observables at a time \cite{SAN}. 
In order to exploit this
effect, we introduce a third observable, 
\mbox{$\hat Z\equiv \sigma^{(3)}_{a}\otimes \sigma^{(3)}_{b}$} and
derive a separability condition which is independent
from Eq.~(\ref{sepent1}).
On one hand, since $\hat Z$ commutes with 
the operators $\hat X$, $\hat Y$ of Eq. (\ref{es1}), 
for a generic state $\rho$ we have
\begin{eqnarray}
H({\mathbb X}, \rho) + H({\mathbb Y}, \rho) + 
H({\mathbb Z}, \rho) \geqslant  0 \; 
\label{sepent2},
\end{eqnarray}
where the equality is achieved on the Bell states of Eq.~(\ref{PBELL}).
On the other hand, when the quantity on the left-hand-side of 
Eq.~(\ref{sepent2}) is evaluated a separable state
it obeys to the following inequality (see App. \ref{PAPPA1}
for the derivation),
\begin{eqnarray}
H({\mathbb X}, \rho_{sep}) + H({\mathbb Y}, \rho_{sep}) + 
H({\mathbb Z}, \rho_{sep}) \geqslant  2 \ln2 \; 
\label{sepent3}.
\end{eqnarray}
This relation provides a weaker separability condition, i.e. a
more sensitive entanglement criteria, than 
Eq. (\ref{sepent1}). In fact, since each of the operators
$\hat X$, $\hat Y$ and $\hat Z$ has only two orthogonal eigenspaces,
the quantities $H({\mathbb X}, \rho_{sep})$, 
$H({\mathbb Y}, \rho_{sep})$ and
$H({\mathbb Z}, \rho_{sep})$ are always smaller than $\ln 2$. Using
this property it is straightforward to show that Eq. (\ref{sepent3})
implies  Eq. (\ref{sepent1}). This means that if a state $\rho$ 
is entangled according to Eq. (\ref{sepent1}) (i.e. if
$\rho$ violates such inequality), then it is also
entangled according to  Eq. (\ref{sepent3}) (i.e. 
$\rho$ violates also this inequality). The opposite, however, is not
true: entangled states that satisfy the inequality
(\ref{sepent1}) but not the inequality (\ref{sepent3}) exist 
(e.g. see the case discussed in Sec. \ref{SezDis}).


\subsection{Operators diagonal in the Bell basis}\label{PGENERAL}

We have seen that the operators
$\hat X$, $\hat Y$ and $\hat Z$ form a
{\em filtering system},
which is partially permeable to entangled states, but which
does not allow any separable states to pass without paying a $2 \ln2$ fee
in uncertainty.
A better insight on this property
can be obtained by
analyzing the decomposition of these observables 
in the Bell basis~(\ref{PBELL}), 
\begin{eqnarray}
\hat X &\equiv& |\Psi_1\rangle\langle \Psi_1| +  |\Psi_2\rangle\langle \Psi_2|
- |\Psi_3\rangle\langle \Psi_3|- |\Psi_4\rangle\langle \Psi_4|\nonumber\\
\hat Y &\equiv& -|\Psi_1\rangle\langle \Psi_1| +  |\Psi_2\rangle\langle \Psi_2|
+ |\Psi_3\rangle\langle \Psi_3|- |\Psi_4\rangle\langle \Psi_4|\nonumber\\
\hat Z &\equiv& |\Psi_1\rangle\langle \Psi_1| -  |\Psi_2\rangle\langle \Psi_2|
+ |\Psi_3\rangle\langle \Psi_3|- |\Psi_4\rangle\langle \Psi_4|\nonumber\;. \\
\label{PPOI}
\end{eqnarray}
Each of the above operators identifies two couples
of Bell states associated with the $+1$ and $-1$
eigenvalue respectively.
By measuring $\hat X$, for instance, we can distinguish 
the first two Bell states from the last two, but we cannot
distinguish  $|\Psi_1\rangle$ from  $|\Psi_2\rangle$ or  
$|\Psi_3\rangle$ from  $|\Psi_4\rangle$.
Moreover $\hat X$, $\hat Y$ and
$\hat Z$ identify different subsets of the Bell basis and
hence form a set of {\em topologically} distinguishable observables.

Consider now the separable state $|00\rangle=(|\Psi_1\rangle
+|\Psi_2 \rangle)/\sqrt{2}$.
On one hand, since this is an eigenstate of $\hat X$, 
the uncertainty of such observable nullifies.
On the other hand, since $|00\rangle$ is a uniform superposition
of distinguishable eigenstates of $\hat Y$,
the measurement of this observable 
gives  $+1$ or $-1$ with probability $1/2$ and produces hence one 
bit ($\ln2$) of entropic uncertainty as requested by the
inequality~(\ref{sepent1}). The same is true also 
for $\hat Z$: the measurement of this observable 
on $|00\rangle$ produces  one bit of uncertainty
in agreement with Eq.~(\ref{sepent3}).

One way to generalize the above result is to design different {\em filtering 
strategies} by selecting appropriate sets of nonlocal operators.
In Table \ref{table1}, we present two examples
that employ observables diagonal in the Bell basis~(\ref{PBELL}).
\begin{table}
\begin{tabular}{c|c|c|c|c}
&$|\Psi_1\rangle$ & $|\Psi_2\rangle$ &$|\Psi_3\rangle$& $|\Psi_4\rangle$
\\\hline 
${\hat X}_1^{(1,3)}$ & $+1$& $-1$ &$-1$&$-1$\\\hline
${\hat X}_2^{(1,3)}$&$-1$&$+1$&$-1$&$-1$\\\hline
${\hat X}_3^{(1,3)}$ &$-1$&$-1$&$+1$&$-1$\\\hline 
${\hat X}_4^{(1,3)}$&$-1$&$-1$&$-1$&$+1$\\ \end{tabular}
\hspace{.2cm}
\begin{tabular}{r|r|r|r|r}
&$|\Psi_1\rangle$ & $|\Psi_2\rangle$ &$|\Psi_3\rangle$
& $|\Psi_4\rangle$\\\hline 
${\hat X}_1^{(1,1,2)}$ & $0$& $0$ &$+1$&$-1$\\\hline
${\hat X}_2^{(1,1,2)}$&$0$&$+1$&$0$&$-1$\\\hline
${\hat X}_3^{(1,1,2)}$ &$0$&$+1$&$-1$&$0$\\\hline 
${\hat X}_4^{(1,1,2)}$&$+1$&$0$&$0$&$-1$\\\hline
${\hat X}_5^{(1,1,2)}$ &$+1$&$0$&$-1$&$0$\\\hline 
${\hat X}_6^{(1,1,2)}$&$+1$&$-1$&$0$&$0$\\
\end{tabular}
\caption{On the left: 
Spectral decomposition of a set of operators ${\hat X}_j^{(1,3)}$ with
$j=1,\cdots,4$ that
distinguish only one of the Bell states of Eq.~(\ref{PBELL}) from
the other three.
Notice that there are at most four of these observables that differ
topologically from each other.
On the right: Spectral decomposition of a set of operators 
${\hat X}_j^{(1,1,2)}$ with $j=1,\cdots,6$
that divide the Bell state into a 
group of two and two groups of one states. There are at most six
of them that are topologically not equivalent.
In both the tables, the value in the $j$th row and $v$th column
is the eigenvalue of the $j$th operator associated with $|\Psi_v\rangle$.
\label{table1}}
\end{table}
In the first case, we use the 
operators ${\hat X}_j^{(1,3)}$ that 
individually distinguish one particular Bell 
state from the remaining three, but
not these last from each other.
As in the case of the operators $\hat X$, $\hat Y$ and $\hat Z$, 
the Bell states~(\ref{PBELL}) are the only ones that can
nullify the entropic uncertainties of all the ${\hat X}_j^{(1,3)}$.
One can use these operators to derive
a separable condition analogous to~(\ref{sepent3}). 
In fact, consider again the state
$|00\rangle$: in this case it diagonalizes  
${\hat X}_3^{(1,3)}$ and ${\hat X}_4^{(1,3)}$ but, according
to Table~\ref{table1}, produces one bit
of uncertainty when measured with
${\hat X}_1^{(1,3)}$ or ${\hat X}_2^{(1,3)}$.
As discussed in the App. \ref{PAPPA2} this result can be formalized
by showing that for any separable state $\rho_{sep}$ one has
\begin{eqnarray}
\sum_{j=1}^4H({\mathbb X}_j^{(1,3)}, \rho_{sep}) \geqslant 2 \ln 2 \; .
\label{Psepent3}
\end{eqnarray}
Analogously to the case of the operators (\ref{PPOI}), 
also here it is possible to
identify a minimum number of elements necessary to produce
a nontrivial minimal uncertainty for separable states.
In the case of the operators $\hat X$, $\hat Y$ and $\hat Z$
such number is two: in fact, there are separable states
that diagonalize at least one of these observables
(see the example of $|00\rangle$ discussed above) 
but there are no separable states that diagonalize  two of them
(see Eq. (\ref{sepent1})).
In the case of the operators ${\hat X}_j^{(1,3)}$ 
such a minimum
number is three since there are examples of separable states
that diagonalize any two of them.

The same considerations applies also to the operators 
$\hat X_j^{(1,1,2)}$ of Table \ref{table1} which divide the 
Bell basis into two subgroups of one state and one subgroup of
two states. In this case we have (see App. \ref{PAPPA3})  
\begin{eqnarray}
\sum_{j=1}^6H({\mathbb X}_j^{(1,1,2)}, \rho_{sep}) \geqslant 5 \ln 2 \; ,
\label{Psepent4}
\end{eqnarray}
since separable states can diagonalize only one
of the $\hat X_j^{(1,1,2)}$ at the time.

The final example we consider here is given
by one single observable which assigns a different eigenvalue
to each of the four Bell states, e.g.
\begin{eqnarray}
\hat X^{(1,1,1,1)}&\equiv& |\Psi_1\rangle\langle \Psi_1| +  
2|\Psi_2\rangle\langle \Psi_2| \nonumber\\
&& + 3|\Psi_3\rangle\langle \Psi_3|+4|\Psi_4\rangle\langle \Psi_4|\;.
\label{PEXTREME} 
\end{eqnarray}
To form a separable state we need to superimpose at least two
different vectors of the Bell basis (see App. \ref{PAPPA4}).
Hence, the measurement of ${\hat X}^{(1,1,1,1)}$
on $\rho_{sep}$ produces at least one bit of uncertainty, i.e.
\begin{eqnarray}
H({\mathbb X}^{(1,1,1,1)}, \rho_{sep}) \geqslant \ln 2 \; .
\label{Psepent5}
\end{eqnarray}
This result is analytically proved in App. \ref{PAPPA4}
and, analogously to Eqs. (\ref{Psepent3}) and (\ref{Psepent4}),
gives a separability condition for a two-qubit system.
In Sec. \ref{s:Phigh} we will show how to generalize the above
inequalities to bipartite  systems
of higher dimension.

\subsection{Noncommuting observables}\label{SezJointSum}

An alternative example of entropic 
separability conditions for a two-qubit system, 
is provided by the three orthogonal components
of the total spin of the system, 
\begin{eqnarray}
{\hat S}_1 &\equiv& \sigma^{(1)}_{a}\otimes \openone_{b} + \openone_{a}\otimes
\sigma^{(1)}_{b}  \nonumber \\
{\hat S}_2 &\equiv& \sigma^{(2)}_{a}\otimes \openone_{b} + \openone_{a}\otimes
\sigma^{(2)}_{b}  \nonumber \\
{\hat S}_3 &\equiv& \sigma^{(3)}_{a}\otimes \openone_{b} + \openone_{a}\otimes
\sigma^{(3)}_{b} 
\;, \label{somma}
\end{eqnarray}
where $\openone_{s}$ is the identity operator on the qubit $s$.
These observables do not commute with each other
but the sum of their entropic uncertainties is nullified by 
$|\Psi_4\rangle$ of (\ref{PBELL}) (this is the singlet state which is an
eigenvector of the total spin). 
In other words, for $\rho$ generic we have again  
\begin{eqnarray}
\sum_{j=1}^{3}H({\mathbb S}_j, \rho) \geqslant  0 \; 
\label{sepent44}.
\end{eqnarray}
As in the previous cases, however, if the state $\rho$ is separable,
a nontrivial constraint on the sum of the entropic uncertainties applies.
This happens because the three components of the total spin do not
share any common separable eigenstate.
In particular in this case one can verify that 
for any $\rho_{sep}$ separable (see App. \ref{PAPPA5}),
\begin{eqnarray}
\sum_{j=1}^{3}H({\mathbb S}_j, \rho_{sep}) \geqslant  3 \ln2 \; 
\label{sepent4}.
\end{eqnarray}


\subsection{Comparison with existing entanglement criteria}\label{SezDis}

Here we analyze the 
{\em sensitivity} in detecting the presence of entanglement
of the inequalities derived in the previous sections.
Following the suggestion of \cite{HOF} we consider the set of Werner states 
\begin{eqnarray}
w_p \equiv  \frac{1-p}{4} \, \openone_a \otimes \openone_b 
\, + \, p \, |\Psi_4\rangle \langle \Psi_4| \; 
\label{werner},
\end{eqnarray}
where $|\Psi_4\rangle$ is the state defined in~(\ref{PBELL})
 and $p\in[0,1]$.
The density matrices $w_p$ are separable if and
only if $p \leqslant 1/3$ \cite{WER}.
To test the relations (\ref{sepent1}) and (\ref{sepent3}) we need to
evaluate the entropic uncertainties defined in Eq.~(\ref{entro}) 
for each one of the
operators $\sigma^{(j)}_{a}\otimes \sigma^{(j)}_{b}$, with $j=1,2,3$.
However, since the states $w_p$ are rotationally invariant,
these quantities are identical 
and it is sufficient to evaluate only one of them.
In particular consider $\hat X = \sigma^{(1)}_{a}\otimes
\sigma^{(1)}_{b}$. 
The projector operators in the eigenspaces of this observable are
$\mathbb X_{+1} \equiv |00\rangle \langle 00| + |11\rangle \langle 11|$
and $\mathbb X_{-1} \equiv |01\rangle \langle 01| + |10\rangle \langle
10|$, so that the probabilities $P_k$ of Eq.~(\ref{entro})
for the state  $w_p$ are $P_{\pm} = (1 \pm
p)/2$. Consequently, we have 
\begin{eqnarray}
&&\!\!H({\mathbb X}, w_p) 
+ H({\mathbb Y}, w_p) =  2 \, {\cal H}_2\left(\frac{1+p}2\right) \; 
\label{werner1} \\
&&\!\!H({\mathbb X}, w_p) + H({\mathbb Y}, w_p) + 
H({\mathbb Z}, w_p) = \nonumber
3 \, {\cal H}_2\left(\frac{1+p}2\right) \;,\\
\label{werner2}
\end{eqnarray} 
with ${\cal H}_2(x)$ the binary entropy function (\ref{PBINARY}).
According to our analysis we can conclude that the state $w_p$ 
is entangled if Eq.~(\ref{werner1}) or (\ref{werner2})
violates the lower bounds established by Eqs. (\ref{sepent1}), 
(\ref{sepent3}), respectively. In the first case,
 this happens for $p>0.78$, while in the second case it is sufficient 
to have $p>0.65$. [This is in agreement with the fact Eq.~(\ref{sepent3})
is a weaker separability condition than Eq.~(\ref{sepent1})].
The same analysis can be repeated for the 
separability conditions (\ref{Psepent3}), (\ref{Psepent4})
and (\ref{Psepent5}) yielding respectively, 
\begin{eqnarray}
&&3{\cal H}_2\left(\frac{1-p}{4}\right) + {\cal H}_2
\left(\frac{1+3p}{4}\right) \geqslant 2\ln 2 \nonumber \\
&&3{\cal H}_2\left(\frac{1+p}{2}\right) +3 {\cal F}
\left(\frac{1+3p}{4}\right) + 9  {\cal F}
\left(\frac{1-p}{4}\right) \geqslant 5\ln 2 \nonumber \\
&& 3{\cal F}\left(\frac{1-p}{4}\right) 
+{\cal F}\left(\frac{1+3p}{4}\right) \geqslant \ln 2 \;, \label{Pw3}
\end{eqnarray}
where ${\cal F}(x)=-x \ln x $.
The fist of these relations is violated for $p>0.68$, the second
for $p>0.72$ and the last for $p>0.74$. Hence
the separability condition (\ref{sepent3}) still provides the
best criterion. However, we cannot
conclude that (\ref{sepent3}) is, in general,
a more sensitive criterion than (\ref{Psepent3}),
(\ref{Psepent4}) and (\ref{Psepent5}). As a matter of fact, there is not
a clear general ordering between these criteria: the best performance of
(\ref{sepent3}) in this case is due to the symmetry of
the states (\ref{werner}). 

Let us now analyze the noncommuting set of observables
analyzed in Sec. ~(\ref{SezJointSum}). Such operators
have only $|\Psi_4\rangle$ as common eigenvector
and hence we expect the separability
condition (\ref{sepent4}) to be better suited to detect
 the presence of entanglement in $w_p$. 
In this case the entropic uncertainty is
\begin{eqnarray}
\sum_{j=1}^3H({\mathbb S}_j, w_p)= 
3  {\cal F}\left(\frac{1+p}2\right)
+ 6  {\cal F}\left(\frac{1-p}4\right) \;,
\label{werner3}
\end{eqnarray}
with ${\cal F}(x)$ defined as in Eq.~(\ref{Pw3}).
By comparing this function with the separability condition
(\ref{sepent4}), we can establish that $w_p$
is entangled for $p > 0.55$.
As expected the entanglement criterion based on this relation is hence 
able to recognize more entangled states than the previous ones.
Nevertheless, our best 
example is still not able to pinpoint the threshold $p=1/3$ as, instead,
the criterion proposed in \cite{HOF} does, or even achieve the threshold
$p=1/2$ as the criterion proposed in \cite{CAM,NOTA2}.
Of course this can be just a consequence of the choice of
the operators we have assumed in developing the
entropic separability conditions proposed here: 
refining this choice
the method might achieve better performances.


\section{Separability conditions for $d\times d$ bipartite systems.}
\label{s:Phigh}

In this section we discuss the entropic separability conditions
for $d \times d$ bipartite systems $\cal S$
where the two 
subsystems $\cal A$ and $\cal B$ are associated with Hilbert spaces of 
equal dimension $d\geqslant 2$.
The situation is not different, in principle, from the qubit case 
analyzed in Sec. \ref{SezSep}.
Also here one needs to define a collection
 of nonlocal operators
$\{ \hat{X}_j \}$ of $\cal S$
 which do not have any common separable eigenvector.
For at least some of these sets one expects the global entropic uncertainty
to have a nontrivial lower bound when
evaluated over the states $\rho_{sep}$ of Eq. (\ref{sepone}).
In other words, define 
\begin{eqnarray}
E &\equiv& \min_{\rho}\sum_{j=1}^{J}H({\mathbb X}_j, \rho)
\label{DDUE}\\		
E_{sep}&\equiv& 
\min_{\rho_{sep}}\sum_{j=1}^{J}H({\mathbb X}_j, \rho_{sep})\;,
\label{DUNO}
\end{eqnarray}
where the first minimization is performed for all $\rho$ of $\cal S$
and the second  only on the separable states (\ref{sepone}) of the system.
$E_{sep}$ is clearly greater or equal to $E$. Here, however, we
are looking for sets of operators 
that have  a {\em strictly positive} gap $\Delta$
between these quantities, i.e.
\begin{eqnarray}
\Delta \equiv E_{sep} - E >0
\label{DZERO}\;.
\end{eqnarray}
For any $\{ {\hat X}_j \}$ for which this inequality
holds we have a separability condition on $\cal S$.
In general the calculation of $\Delta$ is nontrivial 
since it requires to find
global minima of entropic quantities \cite{SHOR}.
In the following section we will present a partial solution of
this problem by
focusing  on a special class of commuting operators ${\hat X}_j$ 
which are diagonal in a maximally entangled basis of $\cal S$. 

\subsection{Bell sets}\label{S:PBELL}

A maximally entangled state $|\Psi_{bell}\rangle$ 
of $\cal S$ is a state that has 
maximally mixed reduced density matrix, e.g.
\begin{eqnarray}
\rho_a&\equiv& \mbox{Tr}_b 
\left [|\Psi_{bell}\rangle \langle\Psi_{bell}|\right] 
= \openone_a /d 
\label{DMAXIMALLY}\;,
\end{eqnarray}
where $\mbox{Tr}_a[\cdots]$ is the
partial trace over the subsystem $\cal A$.
The Bell states (\ref{PBELL}) are an example of 
maximally entangled states for a two-qubit system.
Consider  an orthonormal basis 
$\{ |\Psi_{v}\rangle\}$ composed by these particular vectors.
As in Sec. \ref{PGENERAL} we will consider sets of Bell operators,
i.e. ensembles of nonlocal observables which are  diagonal with 
respect to $\{ |\Psi_v\rangle \}$
and divide the elements of such basis into subgroups of degenerate
eigenvectors. 
For instance, the Bell set $\{ {\hat X}_j^{(M_1,\cdots,M_K)} \}$
contains all the topologically distinguishable observables which
have $K\leqslant d^2$ distinct eigenvalues of degeneracies $M_1, \cdots, M_K$.
Clearly the absolute minimum $E$ of Eq.~(\ref{DDUE}) of each of 
these ensembles is zero (their operators commutes): 
thus, to establish if a Bell set can
be used to derive a separable condition it is sufficient to
evaluate the quantity $E_{sep}$ of Eq.~(\ref{DUNO}).
This problem is still difficult to solve, 
even though in many case it is possible to 
heuristically {\em guess} the right answer.  
In the following we will analyze the simplest 
of these Bell sets which are
respectively the generalization to the case $d\geqslant 2$ 
of the operators ${\hat X}^{(1,1,1,1)}$ and ${\hat X}_j^{(1,3)}$ of 
Sec. (\ref{PGENERAL}): at least in these cases it is
relatively easy to derive analytically the solution.

\subsubsection{First case}
Consider the nonlocal operator
\begin{eqnarray}
\hat X^{(1,\cdots,1)} \equiv  \sum_{v=1}^{d^2} \; v \; |\Psi_{v}\rangle
\langle \Psi_v |\;.
\label{DEXTREME}
\end{eqnarray}
This is the simplest example of a Bell set: it contains a single
element that  distinguishes all the elements of the maximally entangled
basis from each other. 
As shown in App. \ref{PAPPB}, the projection  probability of any
separable state $\rho_{sep}$
on a generic vector of the maximally entangled basis
cannot be greater than $1/d$, i.e.
\begin{eqnarray}
Q_v\equiv \langle \Psi_v | \rho_{sep} | \Psi_v \rangle\leqslant 1/d
\label{DQ}
\end{eqnarray} 
for any $v=1,\cdots, d^2$. 
Hence from the definitions~(\ref{entro}) and~(\ref{DEXTREME}) we have,
\begin{eqnarray}
H(\mathbb{X}^{(1,\cdots,1)}, \rho_{sep})\equiv
-\sum_{v=1}^{d^2} Q_v \ln Q_v 
\geqslant \ln d \label{DCRI1}\;.
\end{eqnarray}
This, together with the fact that $E=0$, shows that the gap $\Delta$ 
of Eq.~(\ref{DZERO})  is $>0$.  
Thus, Eq.~(\ref{DCRI1}) provides an entanglement criterion for
$d \times d$ bipartite systems. For $d=2$ it reduces to
Eq.~(\ref{Psepent5}).

\subsubsection{Second case}
Consider now the Bell set composed by the following 
$d^2$ topologically distinct operators 
\begin{eqnarray}
{\hat X}_v^{(1,d^2-1)}\equiv 2 |\Psi_v\rangle \langle \Psi_v| -\openone 
\label{DCASO2}
\end{eqnarray}
where $v=1,\cdots, d^2$ and 
$\openone$ is the identity operator of $\cal S$.
These observables are the generalization for $d\geqslant 2$ of the operators
${\hat X}_j^{(1,3)}$ defined in Table \ref{table1}: 
each of them divides the maximally entangled basis $\{ |\Psi_v \rangle
\}$ into two groups,
the first containing the eigenvector associated with the 
eigenvalue $+1$ and the second containing the  $d^2-1$ eigenvectors
associated with the eigenvalue $-1$.
We expect  $E_{sep}$ of Eq. (\ref{DUNO}) to
be strictly positive. Roughly speaking, since any separable
state $|\Psi_{sep}\rangle$ must be a superposition of at least
$d$ orthogonal maximally entangled states (see Eq. (\ref{DORA}) below),
one should have an average uncertainty of ${\cal H}_2(1/d)$ from
every ${\hat X}_j^{(1,d^2-1)}$ not diagonalized by $|\Psi_{sep}\rangle$.
Moreover, the same argument can be used to show that 
the minimum number of non diagonalized operators must of the order of
$d$. [In fact, given a collection of $d$ basis elements,
 the maximum number of ${\hat X}_j^{(1,d^2-1)}$ that
assign the  eigenvalue $+1$  to one of such vectors is $d$].
In other words, it is reasonable to assume that the following inequality
applies to any separable state $\rho_{sep}$ of the system,
\begin{eqnarray}
\sum_{j=1}^{d^2} H(\mathbb{X}_j^{(1,d^2-1)}, \rho_{sep})
\geqslant d \;{\cal H}_2(1/d)\label{DCRI2}\;,
\end{eqnarray}
where 
 ${\cal H}_2$ is the binary entropy function of Eq. (\ref{PBINARY}).
[Notice that for $d=2$ the above expression
reduces to the inequality~(\ref{Psepent3})].
To prove Eq. (\ref{DCRI2}) we first notice that 
the entropic uncertainty of the operator $\hat X_j^{(1,d^2-1)}$
can be expressed as
\begin{eqnarray}
H(\mathbb{X}_j^{(1,d^2-1)}, \rho_{sep}) &=&{\cal H}_2(Q_j)
\label{DORABELLA}\;.
\end{eqnarray}
Moreover, given a separable state $\rho_{sep}$, there must be at least
$d$ elements of the maximally entangled basis $\{|\Psi_v\rangle\}$
that have projection probabilities $Q_v>0$. 
In fact, let $r$ be the number of such elements,
then from the normalization condition of $Q_v$ and from the property
(\ref{DQ}) we have
\begin{eqnarray}
1=\sum_{v=1}^{d^2}Q_v \leqslant r /d \; \label{DORA}\;.
\end{eqnarray}
Ordering the $Q_v$ in decreasing order, we can hence 
write the entropic uncertainty of this set of operators
as follows
\begin{eqnarray}
\sum_{j=1}^{d^2} H(\mathbb{X}_j^{(1,d^2-1)}, 
\rho_{sep}) = \sum_{j=1}^{d}
{\cal H}_2(Q_j) +   \sum_{j=d+1}^{d^2}
{\cal H}_2(Q_j)\nonumber \\
 \label{DORABELLA1}\;.
\end{eqnarray}
Every $Q_j$ in the first summation is bigger than the $Q_j$ in the
second summation. We can hence use the property {\bf iii)}
 of the binary entropy given in App. \ref{appA} to derive a lower bound
for~(\ref{DORABELLA1}). Namely, according to Eq.~(\ref{PPROP})  
some positive quantities can be subtracted from each of 
the arguments of the binary 
entropies of the second summation and added to the 
arguments of the binary entropy of the first summation. 
Equations~(\ref{DQ}) and (\ref{DORA}) guarantee that these quantities
can be chosen to transform into $1/d$ all the arguments in the first summation,
nullifying in the meantime all the arguments in the second one.
This concludes the proof because we have generated 
a lower bound of (\ref{DORABELLA1})
made of $d$ binary entropies evaluated in~$1/d$.


\section{Conclusions} \label{SezCon}
The inequalities presented in this paper
are examples of how one can construct separability 
conditions for bipartite systems starting
from entropic uncertainty relations. They
derive from the fact that the common eigenstates of some
nonlocal operators are entangled. 
As in the case of separability conditions obtained from
the Heisenberg-like uncertainty relation 
\cite{REID1,DUA,CAM,KOR,HOF},
this property can be exploited 
to impose bounds on the minimum indetermination
that can be achieved when measuring 
simultaneously these operators
on separable states. 
A detailed analysis of this procedure has been 
provided in the case of two-qubit systems.
The main problem in deriving entropic
separability conditions resides with the fact
that one has to solve a (constrained) minimization
of a concave function (the Shannon entropy). This
in general is a difficult task even in the case of
a commuting set of observables. A possible solution of
this problem could be obtained, for instance,
 by replacing the Shannon
entropy with the R\'enyi entropies which are sometime
easier to handle. At this stage of the investigation it is
not yet clear if entropic separability conditions
are competitive with respect to other strategies (when
they are available): in the examples discussed 
in the paper we have found evidences of the contrary. 
However, since the method proposed here
relays on measurable quantities (the probabilities
of finding the state in the eigenspaces 
associated with certain operators) it can
be a useful tool in characterizing the 
presence of entanglement in many experimental
context.


\appendix
\section{Derivation of the separability conditions for
qubits}\label{appA}
In this Appendix we prove the inequalities
introduced in Sec. \ref{SezSep}
which provide the separability
conditions for the two-qubit system.

In the derivation we will use of the following 
properties of the
binary entropy of Eq.~(\ref{PBINARY}):
\begin{itemize}
\item[{\bf i)}] For $x\in[0,1]$ the function ${\cal H}_2(x)$ is
decreasing in $|1-2x|$, \cite{COVER}.
\item[{\bf ii)}] For $x,y\in [0,1]$ and $x+y\leqslant1$ 
one has 
\begin{eqnarray}
{\cal H}_2(x) + {\cal H}_2(y) \geqslant {\cal H}_2(x+y) 
\label{PPROP1}\;.
\end{eqnarray}
This can be seen by studying the dependence on $y$ of
 the difference between the left-hand-side
and the right-hand-side of Eq. (\ref{PPROP1}).
\item[{\bf iii)}]The previous relation can be 
generalized to show that 
for any $x$, $y$ and $z\in [0,1]$ 
with $1-y\geqslant x\geqslant y\geqslant z$,
we have \begin{eqnarray}
{\cal H}_2(x) + {\cal H}_2(y) \geqslant {\cal H}_2(x+z) + 
{\cal H}_2(y-z)
\label{PPROP}\;.
\end{eqnarray} 
For $z=0$ this is a trivial identity. For $z$ positive the 
above inequality can be obtained by observing that
the first derivative in $z$ of the difference between the 
left-hand-side and the right-hand-side
of Eq. (\ref{PPROP}) is positive for all $z$ in the domain.
\end{itemize}

As in the case of Eq.~(\ref{sepent1}),
we will use the concavity of the Shannon entropy \cite{COVER}
to limit the analysis to the pure separable states $|\Psi_{sep}\rangle$
 of Eq. (\ref{PSEP}):
if the inequalities (\ref{sepent3}), (\ref{Psepent3}), (\ref{Psepent4}) 
and (\ref{Psepent5}) 
apply for this  class of separable states, then they 
hold also for any  mixed separable states of the form~(\ref{sepone}).
The projection probabilities 
$Q_v\equiv |\langle \Psi_{sep}|\Psi_v \rangle|^2$ 
of $|\Psi_{sep}\rangle$ on the Bell vectors
of Eq. (\ref{PBELL}) will be expressed as
\begin{eqnarray}
Q_1&=& (q_0+q_1)/2 \qquad Q_3= (1-q_0+q_2)/2\nonumber\\
Q_2&=& (q_0-q_1)/2 \qquad Q_4= (1-q_0-q_2)/2\;,
 \label{PPROB}
\end{eqnarray}
where $q_1=q\cos(\delta + \gamma)$,  $q_2=q\cos(\delta - \gamma)$ and 
\begin{eqnarray}
q_0&=& |\cos\alpha\cos\beta|^{2} +
|\sin\alpha \sin\beta|^{2} \nonumber\\
q&=&2 \cos\alpha\cos\beta \sin\alpha \sin\beta
\label{PAEB}\;.
\end{eqnarray}
These quantities satisfy the relation
$q_0\pm q = \cos^2(\alpha\mp\beta)$ which 
implies 
\begin{eqnarray}
|q_1|,|q_2|\leqslant |q|\leqslant \min[q_0,1-q_0] \;,
\label{PDOPO1}
\end{eqnarray}
and $Q_v\leqslant 1/2$ for all $v=1,\cdots,4$.

\subsection{Derivation of Eq. (\ref{Psepent5})}\label{PAPPA4}

We start from this separability condition because it is
the easiest to derive. From the definitions (\ref{entro}) and 
(\ref{PEXTREME}) we have
\begin{eqnarray}
H({\mathbb X}^{(1,1,1,1)}, |\Psi_{sep}\rangle) &=&
-\sum_{v=1}^4 Q_v \ln Q_v\label{PZERO} \;,
\end{eqnarray}
with $Q_v$ defined in (\ref{PPROB}).
The inequality~(\ref{Psepent5}) then is a consequence of 
$Q_v\leqslant 1/2$, i.e. $-\ln Q_v \geqslant -\ln(1/2)$.

\subsection{Derivation of Eq. (\ref{sepent3})}\label{PAPPA1}

The spectral decompositions (\ref{PPOI}) allow us
to express the entropic uncertainties of the operators $\hat X$,
$\hat Y$ and $\hat Z$ on the state $|\Psi_{sep}\rangle$
in terms of the probabilities 
$Q_v$ of Eq.~(\ref{PPROB}), i.e. 
\begin{eqnarray}
&&\sum_{j=1}^3H({\mathbb X}_j^{(2,2)}, |\Psi_{sep}\rangle)\;=\;
{\cal H}_2(q_0)\label{PPRIMA} \\
&&\quad+ {\cal H}_2\left(\frac{1+q_2-q_1}{2}\right)+
{\cal H}_2\left(\frac{1+q_1+q_2}{2}\right)
\nonumber \; ,
\end{eqnarray}
where, following the notation introduced in Sec. \ref{PGENERAL},
we defined  $\hat X_1^{(2,2)}\equiv \hat X$, $\hat X_2^{(2,2)}\equiv \hat Y$
and  $\hat X_3^{(2,2)}\equiv \hat Z$.
Using the properties {\bf i)} and {\bf ii)} of ${\cal H}_2$ it is 
possible to derive a lower bound for the above expression.
Consider for instance the case  $q_2\geqslant q_1 \geqslant 0$.
According to {\bf ii)} we 
obtain a lower bound for (\ref{PPRIMA}) by subtracting
$(q_2-q_1)/2$ 
from the argument of the second binary entropy term on  the 
right-hand-side of Eq. (\ref{PPRIMA}) and by adding it 
to the argument of the third
binary entropy term, i.e.
\begin{eqnarray}
\label{PDOPO} 
&& {\cal H}_2\left(\frac{1+q_2-q_1}{2}\right)+
{\cal H}_2\left(\frac{1+q_1+q_2}{2}\right) \nonumber \\ &&\qquad \geqslant
 {\cal H}_2\left(1/2\right)+
{\cal H}_2\left(1/2+q_2\right)\;.
\end{eqnarray}
Using the symmetries of Eq. (\ref{PPRIMA}) and ${\cal H}_2(x)$ it is
easy to show that for $q_1$, $q_2$ generic the above  inequality still
applies if we  replace  $q_2$ with $\max(|q_1|,|q_2|)$ 
in the right-hand-side term.
By replacing $q_j$ with
$\min[q_0, 1-q_0]$ in the previous expression,
the property {\bf i)} of
binary entropy and Eq.~(\ref{PDOPO1}) allow then to
establish the following inequality, 
\begin{eqnarray}
\sum_{j=1}^3H({\mathbb X}_j^{(2,2)}, |\Psi_{sep}\rangle) \geqslant
\ln2 + {\cal G}(q_0) 
\label{PPRIMA2}\; ,
\end{eqnarray}
where, for $q_0\in[0,1]$,
\begin{eqnarray}
{\cal G}(q_0) \equiv {\cal H}_2(q_0) \label{PPRIMA3} 
+{\cal H}_2(1/2+\min[q_0,1-q_0])\;.
\end{eqnarray}
The thesis (\ref{sepent3}) is hence a consequence of the fact
${\cal G}(q_0)$ has absolute minimum equal to $\ln2$ for $q_0=1/2$.
[This last property can be verified by studying the
first derivative of ${\cal G}(q_0)$].

\vspace{.2cm}
\begin{figure}[t]
\begin{center}
\epsfxsize=.7\hsize\leavevmode\epsffile{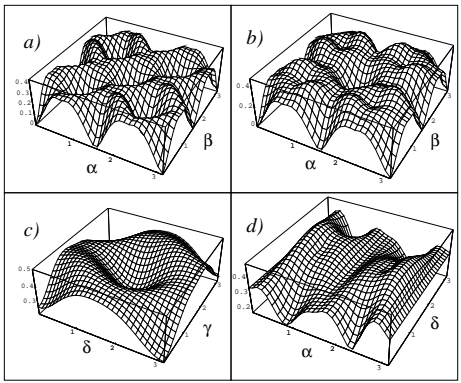}
\end{center}
\caption{Plot of the difference 
between the left-hand-side and the right-hand-side of Eq.~(\ref{sepent4})
for the state $|\Psi_{sep}\rangle$ of Eq.~(\ref{PSEP}): this function
is positive.
In {\em a)} $\delta=\gamma=\pi$;
in {\em b)} $\delta=-\pi/2$, and $\gamma=\pi$; in {\em c)}
$\alpha=-\pi/3$, 
and $\beta=-\pi/5$; in {\em d)} $\beta=\pi/3$, and $\gamma=0$.}
\label{fig2}
\end{figure}

\subsection{Derivation of Eq. (\ref{Psepent3})}\label{PAPPA2}

The derivation of the separability condition~(\ref{Psepent3}) 
proceeds as in the previous section.
First we express the entropic uncertainty relation of
the operators ${\hat X}_j^{(1,3)}$ in terms of the probabilities
$Q_v$ of Eq. (\ref{PPROB}). In this case, according to the
spectral decomposition of Table \ref{table1} we have
\begin{eqnarray}
\sum_{j=1}^4H({\mathbb X}_j^{(1,3)}, |\Psi_{sep}\rangle) 
= \sum_{j=1}^4 {\cal H}_2 (Q_v)\;.
\label{PSECONDA}
\end{eqnarray}
From the definition (\ref{PBINARY})
we can write the first two contribution on the right-hand-side of
Eq. (\ref{PSECONDA}) as
\begin{eqnarray}
&&{\cal H}_2 (Q_1)+{\cal H}_2 (Q_2) = 
-q_0 \ln q_0 -(2-q_0)\ln(2-q_0) \nonumber \\
&& + q_0 \; {\cal H}_2\left(\frac{q_0+q_1}{2q_0}\right) + 
(2-q_0)\; {\cal H}_2\left(\frac{2-q_0 + q_1}{2(2-q_0)}\right) \;,
\nonumber\\
\label{PSECONDA1}
\end{eqnarray}
which, according to the property {\bf i)} of the binary entropy, reaches
its minimum value when $|q_1|$ reaches its maximum $\min[q_0,1-q_0]$. 
Analogously we can show that 
${\cal H}_2 (Q_3)+{\cal H}_2 (Q_4)$ has its minimum for $|q_2|=
\min[q_0,1-q_0]$.
Together these relations provide the following lower 
bound of the total entropic uncertainty~(\ref{PSECONDA})
\begin{eqnarray}
\sum_{j=1}^4H({\mathbb X}_j^{(1,3)}, |\Psi_{sep}\rangle) 
\geqslant \ln2 + {\cal G}(q_0)\;,
\label{PSECONDA2}
\end{eqnarray}
where ${\cal G}(q_0)$ is the function defined in Eq. (\ref{PPRIMA3}).
The thesis (\ref{Psepent3}) follows once again from the fact that
the absolute minimum of ${\cal G}(q_0)$ is $\ln2$.

\subsection{Derivation of Eq. (\ref{Psepent4})}\label{PAPPA3}

The separability condition (\ref{Psepent4}) is slightly
more demanding to prove than the previous ones.
Here, according to Table \ref{table1}, we have six different
observables ${\hat X}_j^{(1,1,2)}$, 
each with three distinct eigenvalues: the 
two times degenerate eigenvalue $0$ and the non degenerate eigenvalues $\pm 1$.
For instance, for $j=1$ we have
\begin{eqnarray}
H({\mathbb X}_1^{(1,1,2)}, |\Psi_{sep}\rangle)\; &=&\;
-(Q_1+Q_2)\ln(Q_1+Q_2) \nonumber \\
&&-Q_3 \ln Q_3 -Q_4\ln Q_4
\label{PTERZA}\;.
\end{eqnarray}
Some elementary simplifications allow to express
the total entropic uncertainty as the sum
of three contributions
\begin{eqnarray}
\sum_{j=1}^6 H({\mathbb X}_1^{(1,1,2)}, |\Psi_{sep}\rangle)\; = \;
f_0+f_1+f_2 
\label{PTERZA2}\;,
\end{eqnarray}
where
\begin{eqnarray}
f_0 &\equiv& {\cal H}_2(q_0) + {\cal H}_2\left(\frac{1+q_2-q_1}{2}\right)+
{\cal H}_2\left(\frac{1+q_1+q_2}{2}\right)\nonumber\\
f_1 &\equiv&3 q_0 {\cal H}_2\left(\frac{q_0+q_1}{2q_0}\right)-3q_0\ln q_0
\label{treffe}  \\
f_2 &\equiv&3 (1-q_0) {\cal H}_2\left(\frac{1-q_0+q_2}{2(1-q_0)}\right)-
3(1-q_0)\ln (1-q_0) \nonumber \;.
\end{eqnarray}
We notice that $f_0$ is the function given in Eq. 
(\ref{PPRIMA}) and that, according to the analysis
of App.  \ref{PAPPA1}, it is always greater than or
equal to $2\ln2$.
On the other hand, using the property {\bf i)} of the 
binary entropy, we observe that $f_1$ and $f_2$ 
are decreasing functions of $|q_1|$ and $|q_2|$, respectively.
Hence using Eq. (\ref{PDOPO1}) we obtain
\begin{eqnarray}
f_1 + f_2 \geqslant 3 \ln 2 + \frac{3}{2} {\cal H}_2(2\min[q_0,1-q_0]) 
\geqslant 3 \ln 2 \nonumber\\
\label{PTERZO3}
\end{eqnarray}
which, together with $f_0\geqslant 2 \ln 2$ proves the thesis (\ref{Psepent4}) 
when replaced into Eq. (\ref{PTERZA2}).

\subsection{Derivation of Eq. (\ref{sepent4})}\label{PAPPA5}

The proof proceeds as the previous ones: we evaluate the left-hand-side of
Eq.~(\ref{sepent4}) on the state $|\Psi_{sep}\rangle$ of Eq.(\ref{PSEP}) and
we look for its minimum value. 
Each of the operators~(\ref{somma}) has three orthogonal eigenspaces
(relative to the non degenerate eigenvalues $\pm 2$ and the two time
degenerate eigenvalue $0$), so that its entropic uncertainties
(\ref{entro}) has three contributions, i.e.
\begin{widetext}
\begin{eqnarray}
&H({\mathbb S}_1, |\Psi_{sep}\rangle) =
 {\cal F}(|\cos\alpha \sin\beta|^{2}+|\sin\alpha \cos\beta|^{2}) 
+{\cal F}(|\cos\alpha \cos\beta|^{2})+ {\cal F}(|\sin\alpha \sin\beta|^{2})&
\nonumber\\
&H({\mathbb S}_2, |\Psi_{sep}\rangle)=
 {\cal F}[(1-\sin\delta \sin\gamma \sin(2\alpha)\sin(2\beta))/2] 
+{\cal F}\{[(1-\sin\delta \sin(2\alpha))(1-\sin\gamma
\sin(2\beta)]/4\}&
\nonumber \\
&+{\cal F}\{[(1+\sin\delta \sin(2\alpha))(1+\sin\gamma \sin(2\beta)]/4\}&
\label{ris1} \\
&H({\mathbb S}_3, |\Psi_{sep}\rangle)=
{\cal F}[(1-\cos\delta \cos\gamma \sin(2\alpha)\sin(2\beta))/2] 
+{\cal F}\{[(1-\cos\delta \sin(2\alpha))(1-\cos\gamma
\sin(2\beta)]/4\}&
\nonumber \\
&+{\cal F}\{[(1+\cos\delta \sin(2\alpha))(1+\cos\gamma
\sin(2\beta)]/4\}
\;, &
\nonumber
\end{eqnarray}
\end{widetext}
with ${\cal F}(x)\equiv -x \ln x$. By summing these terms we obtain
the total entropic uncertainty of the operators ${\hat S}_j$ on the
state $|\Psi_{sep}\rangle$.
An analytical study of the resulting expression is very demanding: however
 a simple numerical analysis shows that the minimum of this
function  is  $3 \ln2$ (see  Fig. \ref{fig2}).
Notice that the inequality (\ref{sepent4}) becomes an identity by choosing
(for instance) $\alpha=0$ and $\beta=\pi/2$, i.e.
$|\Psi_{sep}\rangle=|01\rangle$.

\section{Properties of the maximally entangled states}\label{PAPPB}
In this section we derive some useful relations for the
maximally entangled states introduced in Sec. \ref{s:Phigh}.
\paragraph*{Superposition with separable states:--}
First of all we show that given a maximally entangled state
$|\Psi_{bell}\rangle$ and a generic separable state $\rho_{sep}$ 
of $\cal S$ we have
\begin{eqnarray}
\langle \Psi_{bell} | \rho_{sep} | \Psi_{bell} \rangle \leqslant 
1/d \;,
\label{PAPPB1}
\end{eqnarray}
where $d$ is dimension of the subsystems $\cal A$ and $\cal B$.
Since $\rho_{sep}$ can be expressed as a convex convolution of
pure separable states, to prove Eq. (\ref{PAPPB1})
it is sufficient to show that it holds for any 
$|\Psi_{sep}\rangle=|\psi\rangle_a\otimes|\phi\rangle_b$.
Consider an orthonormal basis $\{ |\psi_i\rangle_a \}$ of $\cal A$ and
an orthonormal basis  $\{ |\phi_l\rangle_b \}$ 
of $\cal B$ (here $i,l=1,\cdots,d$) and construct the
orthonormal basis of $\cal S$ made of the separable 
pure states 
$|\psi_i\rangle_a \otimes |\phi_l\rangle_b\equiv |\psi_i,\phi_l\rangle$.
Expanding $|\Psi_{bell}\rangle$ in this
basis and using the property (\ref{DMAXIMALLY}) one can verify that
the probabilities $Q_{il}\equiv |\langle \Psi_{bell}  
| \psi_i,\phi_i\rangle|^2$ 
satisfy the relations
\begin{eqnarray}
\sum_{i'=1}^{d} Q_{i'l} = \sum_{l'=1}^{d} Q_{il'} = 1/d \;,
\label{PAPPB2}
\end{eqnarray}
for all $i$ and $l$. 
Since $Q_{il}$ are positive quantities, Eq.~(\ref{PAPPB2}) implies that
each of them cannot be greater than~$1/d$. This proves the
thesis (\ref{PAPPB1}) 
since the vectors $|\psi_i\rangle_a$ and $|\phi_l\rangle_b$ can be
chosen arbitrarily.
\paragraph*{Basis of maximally entangled states:--} 
An orthonormal basis of maximally entangled state
can be constructed from the separable basis 
$\{ |\psi_i, \phi_l \rangle\}$
of $\cal S$ introduced in the previous section.
For instance, divide the elements of the separable basis
into $d$ orthogonal subsets, the first containing the states 
$|\psi_i , \phi_{i \oplus 1}\rangle$ with $i=1,\cdots d$
(here $\oplus$ is sum modulus $d$),
the second containing the states  
$|\psi_i, \phi_{i \oplus 2}\rangle$ with $i=1,\cdots d$,
etc. Now from each of these subsets we can extract $d$ orthogonal
maximally entangled state by applying a quantum Fourier 
transform~\cite{CHU} to the subset elements: the orthogonality between  
the subsets guarantees that at the hand we have 
generated $d\times d$ orthogonal maximally entangled states, i.e.
a basis of $\cal S$.



\begin{references}
\bibitem{NOTA1}The standard expression $H({\hat X}, \rho)$ 
for the entropic uncertainty of the operator ${\hat X}$
is given by the Shannon entropy of the 
projection probabilities of $\rho$ over a particular basis of
eigenstates of $\hat X$ \cite{DEU,MAJ,MAA,SAN}. In other words, consider
an orthonormal basis of eigenvectors $\{ |x_{k,m} \rangle\}$
of the operator $\hat X$, 
such that $\hat{X}|x_{k,m}\rangle = x_k |x_{k,m}\rangle$, 
with $x_k$ eigenvalue of degeneracy $M_k$.
One defines 
\begin{eqnarray}
    H({\hat X}, \rho)  \equiv -\sum_{k=1}^{K} \sum_{m=1}^{M_k} P_{k,m} \ln 
P_{k,m}\;,
    \label{Pentro}
\end{eqnarray}
where $P_{k,m}\equiv \langle x_{k,m} | \rho |x_{k,m} \rangle$ 
is the probability
associated with the $k, m$th  basis element.
The quantity (\ref{Pentro}) coincides with  $H({\mathbb X}, \rho)$
of Eq. (\ref{entro}) only if ${\hat X}$ is non degenerate
(i.e. if $M_k=1$ for all $k$), otherwise one can show that 
$H({\hat X}, \rho) \geqslant H({\mathbb X}, \rho)$.
Moreover for degenerate observables, the definition of 
$H({\hat X}, \rho)$ depends on the particular choice 
of the set $\{|x_{k,m} \rangle \}$, 
while $H({\mathbb X}, \rho)$ clearly does not.
\bibitem{NOTTATA}The condition (\ref{sepent1}) does not depend on the choice of
the basis $\{|0\rangle, |1\rangle \}$: we can thus generalize
this relation  to the case where 
$\hat X \equiv \sigma_{a}(\hat n_{a})\otimes \sigma_{b}(\hat n_{b})$,
$\hat Y \equiv \sigma_{a}(\hat n_{a}^{\perp})\otimes \sigma_{b}(\hat 
n_{b}^{\perp})$, $\sigma_{s}(\hat n_{s})$ being the Pauli operator of 
the $s$ qubit 
in the $\hat n_{s}$ direction, and 
$\hat n_{s}^{\perp}$ a unit vector orthogonal to $\hat n_{s}$.
\bibitem{NOTA2} To show that the separability criterion given in 
\cite{CAM} recognizes  the state $w_p$ of Eq.~(\ref{werner}) 
as entangled  only if $p>1/2$, define the operators
\begin{eqnarray}
&&\hat{U}=\sigma_a^{(1)}\otimes \openone_b-\openone_a\otimes
\sigma_b^{(1)}\label{appC1}\\
&&\hat{V}=\sigma_a^{(2)}\otimes \openone_b-\openone_a\otimes
\sigma_b^{(2)}\nonumber\; .
\end{eqnarray}
According to Eq.~(22) of \cite{CAM} any separable state $\rho_{sep}$ must
satisfies the inequality
\begin{eqnarray}
\langle \Delta^{2} \hat{U}\rangle  \langle \Delta^{2} \hat{V}\rangle
\geqslant 4 |\langle  \sigma_a^{(3)} \otimes \sigma_b^{(3)} \rangle |^2
\label{appC2}\;,
\end{eqnarray}
where $\langle \cdots \rangle \equiv \mbox{Tr}[\cdots \rho_{sep}]$
and $\Delta^2\hat{U}\equiv \hat{U}^2 -\langle \hat{U} \rangle^2$.
When applied to the state $w_p$ this inequality gives 
$(1-p)^2 \geqslant p^2$ which is verified for all $p\leqslant 1/2$.
\end{references}
\end{document}